\documentclass[letterpaper]{IEEEtran}
\IEEEoverridecommandlockouts
\usepackage{cite}
\usepackage{amsmath,amssymb,amsfonts}
\usepackage{graphicx}
\usepackage{textcomp}
\usepackage{xcolor}
\def\BibTeX{{\rm B\kern-.05em{\sc i\kern-.025em b}\kern-.08em
    T\kern-.1667em\lower.7ex\hbox{E}\kern-.125emX}}

\DeclareMathOperator*{\argmax}{arg\,max}
\DeclareMathOperator*{\argmin}{arg\,min}
\usepackage{algorithm}
\usepackage{algorithmic}

\makeatletter
\newcommand\fs@norules{\def\@fs@cfont{\bfseries}\let\@fs@capt\floatc@ruled
  \def\@fs@pre{}%
  \def\@fs@post{}%
  \def\@fs@mid{\kern3pt}%
  \let\@fs@iftopcapt\iftrue}
\makeatother
\restylefloat{algorithm}

\begin{document}

\title{Enhancing V2X Communications with UAV-mounted Reconfigurable Intelligent Surfaces \\
\thanks{The presented work has been funded by the National Science Centre in Poland within the project (no. 2021/43/B/ST7/01365) of the OPUS programme.}
}

\author{\IEEEauthorblockN{Salim~Janji}
\IEEEauthorblockA{\textit{Institute of Radiocommunications} \\
\textit{Poznan University of Technology}\\
Poznań, Poland \\
salim.janji@doctorate.put.poznan.pl; ORCID:0000-0002-1460-9250 \\}
\and
\IEEEauthorblockN{Pawe\l~Sroka}
\IEEEauthorblockA{\textit{Institute of Radiocommunications} \\
\textit{Poznan University of Technology}\\
Poznań, Poland \\
pawel.sroka@put.poznan.pl; ORCID:0000-0003-0553-7088 \\}
\and
\IEEEauthorblockN{Adrian Kliks}
\IEEEauthorblockA{\textit{Institute of Radiocommunications} \\
\textit{Poznan University of Technology}\\
Poznań, Poland \\
adrian.kliks@put.poznan.pl; ORCID: 0000-0001-6766-7836 \\}
}

\maketitle
\thispagestyle{empty}
\begin{abstract}
This paper addresses the crucial need for reliable wireless communication in vehicular networks, particularly vital for the safety and efficacy of (semi-)autonomous driving amid increasing traffic. We explore the use of Reconfigurable Intelligent Surfaces (RISes) mounted on Drone Relay Stations (DRS) to enhance communication reliability. Our study formulates an optimization problem to pinpoint the optimal location and orientation of the DRS, thereby creating an additional propagation path for vehicle-to-everything (V2X) communications. We introduce a heuristic approach that combines trajectory optimization for DRS positioning and a Q-learning scheme for RIS orientation. Our results not only confirm the convergence of the Q-learning algorithm but also demonstrate significant communication improvements achieved by integrating a DRS into V2X networks. \footnote{Copyright © 2024 IEEE. Personal use is permitted. For any other purposes, permission must be obtained from the IEEE by emailing pubs-permissions@ieee.org. This is the author’s version of an article that has been published in the proceedings of the
22nd IEEE International Conference on Pervasive Computing and Communications Workshops and other Affiliated Events, PerCom Workshops 2024 by the IEEE. Changes were made to this version by the publisher before publication, the final version of the record is available at: https://dx.doi.org/10.1109/PerComWorkshops59983.2024.10502957. To cite the paper use: S. Janji, P. Sroka and A. Kliks, "Enhancing V2X Communications with UAV-mounted Reconfigurable Intelligent Surfaces," \textit{2024 IEEE International Conference on Pervasive Computing and Communications Workshops and other Affiliated Events (PerCom Workshops)}, Biarritz, France, 2024, pp. 708-713, doi: 10.1109/PerComWorkshops59983.2024.10502957 or visit https://ieeexplore.ieee.org/document/10502957.}
\end{abstract}


\section{Introduction}
\label{sec:intro}


With the continuously increasing proliferation of moving vehicles (cars, drones, bicycles, scooters, etc.), the requirement for guaranteed safe, reliable, yet effective wireless communications become dominant \cite{Ucar2018}. Various technological solutions have been proposed to support Vehicle-to-Vehicle (V2V) or, in a broader sense, Vehicle-to-Everything (V2X) communications schemes. This, in consequence, entails the need for flexible system design and ad-hoc reactions to the changing needs in such a distributed and pervasive communication structure. 
The high density of vehicles also causes the increased data traffic observed in the frequency spectrum. Currently, to exchange information, the communication devices may either use dedicated short-range communications (DSRC) \cite{IEEE80211}, or cellular networks (in the mode called cellular-V2X, C-V2X) \cite{VUKAD2018}. Research is being performed to address those challenges. One of the approaches assumes that the overall amount of spectrum allocated to V2X purposes should be increased \cite{Shi2014}. The authors of \cite{Gill2023} proposed the use of a vehicular dynamic spectrum access scheme to utilize the available band more efficiently. In one of our prior papers \cite{Sroka2022}, we proposed the application of reinforcement learning to support dynamic channel allocation to platoons driving along the highway. 

One separate group of researchers deals with the application of drones (also called unmanned aerial vehicles, UAVs) for supporting the functioning of vehicular networks. For some time, the drone base stations (DBS) or flying base stations constitute the concept investigated in rich literature to improve the effectiveness of ground communication \cite{Mozaffari2019}.

Moreover, the integration of reconfigurable intelligent surfaces (RISes) in wireless communications has been attracting global attention due to their innovative impact on signal propagation \cite{Liu2021}. These nearly passive devices can be installed in various environmental settings like walls and buildings, offering new ways to manipulate signal paths, including directed reflection contrary to classical optical geometry. RISes also provide functionalities such as controlled diffusion, refraction, and absorption. 

This paper explores the enhancement of V2X communications by mounting RISes on unmanned aerial vehicles (UAVs), enabling vehicles to communicate via both direct and RIS-mediated virtual links. We focus on optimizing the location of a drone relay station (DRS) equipped with an RIS, aiming to establish an additional virtual line-of-sight (LOS) path for V2X pairs. Our study encompasses controlling the DRS's trajectory towards the optimal endpoint and its RIS's xy orientation using a reinforcement learning method, with the goal of maximizing V2X communication rates.

The remainder of the paper is arranged as follows. In the next section, we briefly present the system model, including a discussion on the channel model. Next, in Sec. III we outline the considered problem formulation, with the proposed solutions for trajectory planning and orientation control presented in Sec. IV. It is followed by the analysis of the simulation results. Finally, the paper is then concluded. 

\section{System model}
\label{sec:system_model}
\begin{figure}
      \centering
\includegraphics[width=0.48\textwidth]{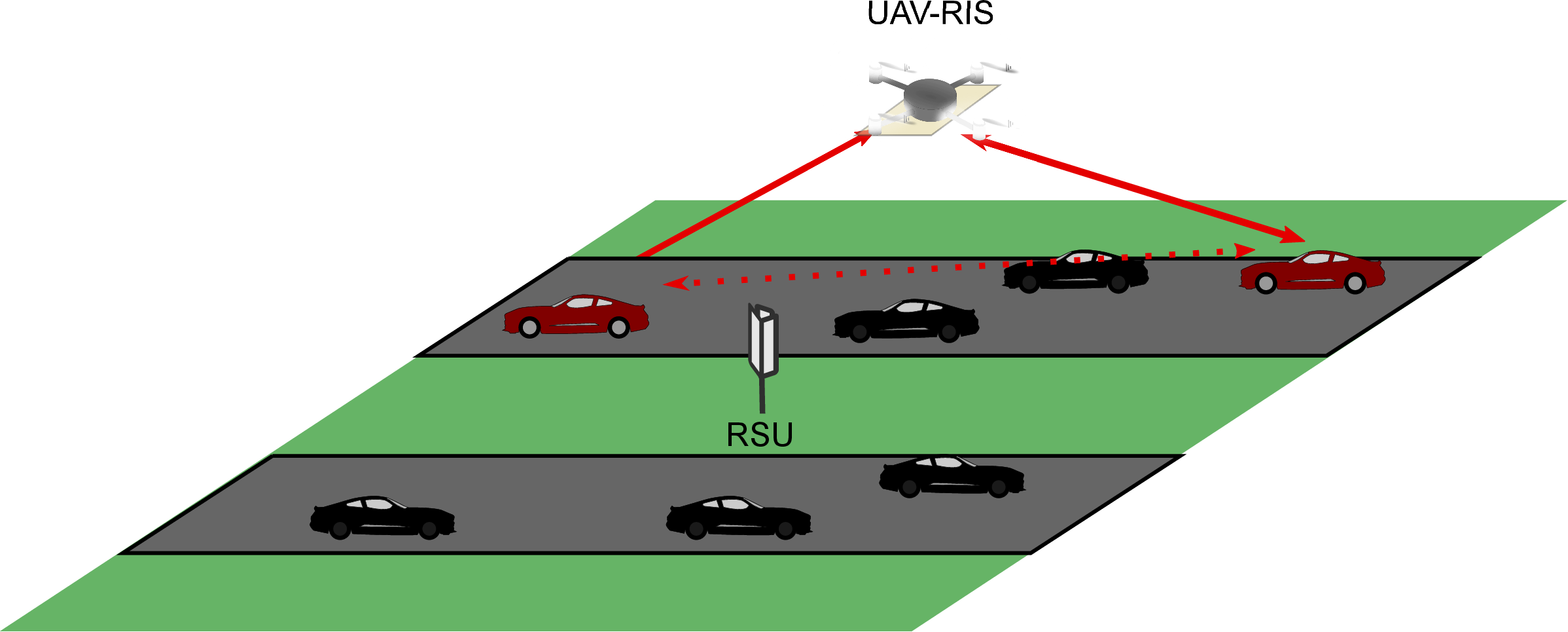}
  \caption{Illustration of the considered scenario with communicating pairs of vehicles moving along a motorway. RIS-equipped UAV is positioned to improve the V2X communications links.}
  \label{fig:scenario}
\end{figure}

In this work, we consider a V2X communications scenario where the connected vehicles move along motorway lanes, as shown in Fig. \ref{fig:scenario}. Two communication types are possible - vehicle-to-vehicle (V2V) transmission, where two cars belonging to the same platoon or fleet are willing to exchange data, and vehicle-to-infrastructure (V2I), where a vehicle is connecting to the roadside unit (RSU) located in the green separation between the lanes. We assume that there are in general $K$ communicating pairs, where for each pair $k$ the location of the involved units (vehicles or RSU)  in 3D at time $t$ is denoted as $(\boldsymbol{L}_{1}^{k}(t),\boldsymbol{L}_{2}^{k}(t))$. Each vehicle or RSU is located at $\boldsymbol{L}_{i}(t) = (x_i(t), y_i(t), h_i(t))$, where $x_i(t)$ represents the lateral position, $y_i(t)$ the longitudinal position, and $h_i(t)$ the height of node $i$. Vehicles move at a constant velocity $v_i$, and their heights are uniformly distributed between 1.5 and 2 meters.

As the distances between the communicating nodes may be significant, the reliability and throughput of the direct V2X links might be insufficient. Thus, we consider the availability of a secondary link, provided by the deployment of a UAV carrying a nearly-passive RIS, denoted also as a DRS, acting as a reflector for V2X transmissions. The location of the UAV is represented with $\boldsymbol{L}_{D}(t) = (x_D(t), y_D(t), h_D(t))$ with similar meaning as for other nodes. We also consider the drone maximum movement velocity of $v_D$, where $\forall_i v_D \geq v_i$. Furthermore, the reflecting capabilities of the UAV-mounted RIS depend on its orientation matrix $\boldsymbol{O}(t)$ with dimensions $3 \times 3$, which defines the orientation of RIS axes.

 We account for the availability of both the direct V2X links and the DRS-assisted links, where both of them might be used jointly to increase the link capacity or reliability. In this study, for the sake of simplicity and to facilitate a clearer analysis of the RIS capabilities, we make the assumption that the virtual link via the RIS and the direct link between the transmitter and receiver are constructively added. For the direct V2V link we assume the propagation model proposed in \cite{Giordani2019} for Highway scenario. However, as we assume that there are no fixed obstacles, such as buildings, there can be only a line-of-sight (LoS) or non-line-of-sight due to vehicles (NLoSv) propagation. In the latter case (NLoSv), the overall path-loss depends, beside the distance between the transmitter and the receiver, also on the height of vehicles in between. In case of V2I communications, on the other hand, we employ the 3GPP Urban Macro (UMa) model to model signal propagation, according to \cite{3gpp38901}.
 
To model the signal propagation in the link via DRS we employ the models proposed for RIS-aided transmission in \cite{Tang2021}. In general two transmission scenarios are possible: far-field or near-field beamforming, selected based on the relation between the distance from the transmitter to RIS $d_{1}$ or RIS to the receiver $d_{2}$ and the Faunhofer distance, calculated as $d_{Fr} = \frac{2D^2}{\lambda}$, where $D$ is the largest dimension of RIS array (e.g. the diagonal) and $\lambda = c/f_c$ is the wavelength \cite{Tang2021}. If $d_{1} > d_{Fr}$ and $d_{2} > d_{Fr}$ far-field beamforming case is assumed, with near-field beamforming considered otherwise. Due to the frequency used and the minimum height of DRS in our scenario, $d_{1}$ and $d_{2}$ are always greater than $d_{Fr}$. Hence, we use only far-field beamforming path-loss formula from \cite{Tang2021}. 

The propagation in the far-field beamforming case depend heavily on the relation between the transmission path axis and the RIS orientation axis, represented with the azimuth angles for both the transmitter $\varphi_t$ and the receiver $\varphi_r$, as well as the elevation angles for both the transmitter $\theta_t$ and the receiver $\theta_r$. The propagation model we use in this case follows the path-loss formula from \cite{Tang2021}:
\begin{equation}
\begin{split}
    PL_{f-f} &=  \frac{64 \pi^3 d_{1}^2 d_{2}^2 }{G_t G_r G M^2 N^2 d_{x} d_{y} \lambda^2 F(\theta_t), F(\theta_r) A^2 \left\vert\Psi\right\vert^2}\\
    \Psi = & \frac{sinc(\frac{M\pi}{\lambda}(\sin{\theta_t}\cos{\varphi_t}+\sin{\theta_r}\cos{\varphi_r})d_x)}{sinc(\frac{\pi}{\lambda}(\sin{\theta_t}\cos{\varphi_t}+\sin{\theta_r}\cos{\varphi_r})d_x)} \\
    &\times \frac{sinc(\frac{N\pi}{\lambda}(\sin{\theta_t}\sin{\varphi_t}+\sin{\theta_r}\sin{\varphi_r})d_x)}{sinc(\frac{\pi}{\lambda}(\sin{\theta_t}\sin{\varphi_t}+\sin{\theta_r}\sin{\varphi_r})d_x)},
\end{split}
    \label{eq:ris_ff_ploss}
\end{equation}
where $G_t$, $G_r$, and $G$ are the transmitter, receiver and RIS antenna element gain, $M$ and $N$ is the number of rows and columns of RIS unit cells (elements), that are distributed uniformly with a spacing of $d_x$ vertically and $d_y$ horizontally. Further, $F(\theta_t)$ and $F(\theta_r)$ are the normalized radiation patterns of RIS unit cells towards the transmitter and the receiver, respectively, given in (\ref{eq:ris_radiation}), while $A$ is the reflection coefficient amplitude for RIS unit cell. We consider the same normalized radiation pattern for all unit cells, depending only on the elevation angle $\theta$ and independent of the azimuth $\varphi$, defined as:
\begin{equation}
    F(\theta) = \begin{cases}  
    \cos^3{\theta} & \theta \in \left[0, \frac{\pi}{2}\right]\\
    0 & \theta \in \left(\frac{\pi}{2}, \pi\right]
\end{cases}
    \label{eq:ris_radiation}
\end{equation}

\begin{figure}[!htb]
\centering
  \includegraphics[width=0.48\textwidth]{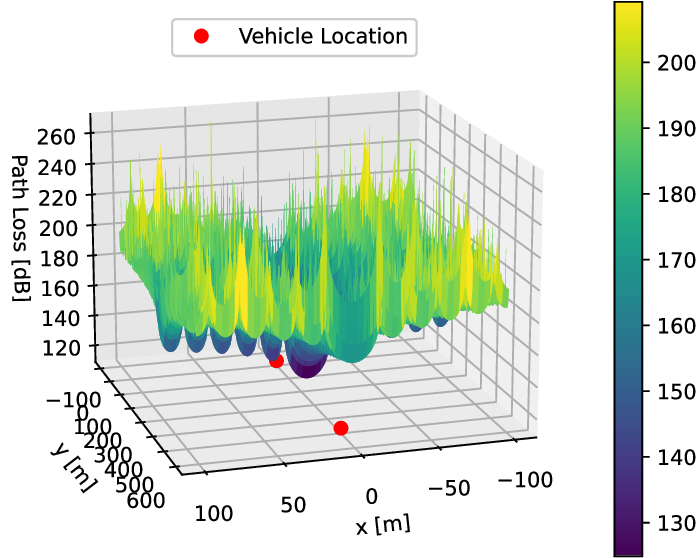}
  \caption{Path loss of the virtual link for different 2D locations of the DRS and $h_D=500$. The V2X pair is marked by the red circles, and the orientation of the DRS is fixed. Notice the high variability of the path loss values due to the sensitivity of the sinc functions to the relations between the azimuth and elevation angles of the link.}
  \label{fig:path_loss_surf}
\end{figure}

An example of the resulting total path-loss for a DRS-assisted link depending on the DRS location with respect to a predefined vehicles pair is shown in Fig.\ref{fig:path_loss_surf}. One can notice the peaks caused by the impact of azimuth and elevation angles.

Based on the calculated path-loss, the throughput of the considered link for pair $k$ can be calculated using the modified Shannon formula \cite{Mogensen2007} given as:
\begin{equation}
\begin{split}
R_{(k)} = \eta B_{\text{eff}}\log_2 (1+SNR_{(k)}),
\end{split}
\end{equation}
where $\eta$ is the bit-stream link effectiveness (fraction of data bits in the total number of bits), $B_{\text{eff}}$ is the normalized effective bandwidth (fraction of used bandwidth to the total bandwidth including guard bands) and $SNR_{(k)}$ is the signal-to-noise ratio (SNR) of the $k$-th pair link calculated as:
\begin{equation}
SNR_{(k)} = \frac{P_t}{PL_{\text{eff}} \sigma_{N}^2},
\end{equation}
where $P_t$ is the transmit power, $PL_{\text{eff}}$ is the effective path-loss and $\sigma_{N}^2$ is the noise variance.
\section{Problem formulation}
\label{sec:problem}
In this work we consider a DRS-aided V2X communication, where an UAV carrying an RIS is providing an additional link supporting direct transmission. Therefore, in order to maximize the gains from using airborne RIS, its location and orientation shall be optimized towards throughput maximization. However, as in general there might be $K$ communicating pairs, the solution to the problem of finding the optimal location and orientation of RIS might not be feasible, as the optimization of certain links might lead to conflicting decisions. Thus, for simplicity of analysis, in the following work we focus on finding the best DRS location and orientation assuming only a single communication link is optimized.

Considering the instantaneous positions of a V2X communication pair, represented as $\boldsymbol{L}_{1}(t)$ and $\boldsymbol{L}_{2}(t)$, along with the position of the DRS, $\boldsymbol{L}_{\text{D}}(t)$, and the 3x3 orientation matrix $\boldsymbol{O}(t)$, the objective is to design the DRS's future trajectory. Specifically, the task is to determine $\boldsymbol{L}_{\text{D}}(t + nT_s)$ for $n \in \{1, 2, \ldots, N\}$, where $T_s$ is the planning time step, and $N$ is the number of steps, determined by the duration of the V2X pair's transmission or their presence within the designated highway segment, which is confined by $(x_{\min}, x_{\max})$ on the $x$-axis and $(y_{\min}, y_{\max})$ on the $y$-axis.
Furthermore, the DRS is capable of modifying the orientation of the RIS along the xy-plane, denoted by $\boldsymbol{O}(t)$, at a rotation rate of $\Gamma_D$ $[\frac{rad}{s}]$. Thus, it is essential to establish not only the trajectory but also the orientation $\boldsymbol{O}^{(n)}$ at each time step. For the sake of clarity and consistency throughout this paper, the time index will be represented as a superscript in all subsequent references. Finally, we assume uniform direction and speed $[\frac{m}{s}]$ within each time step, constrained by the maximum speed $v_D$.

Our objective is to maximize the V2X pair's transmission rate $R^{(n)}$, determined using [RIS MODEL] based on the V2X locations and the DRS's position and orientation. The optimization problem is thus formulated as:

\begin{subequations}\label{eqn:optimization0}
\begin{align}
    &\;\;\;\;\;\;\;\;\;\;\;\;\;\;\;\;\max_{\boldsymbol{O}^{(n)}, \boldsymbol{L}^{(n)}_{\text{D}} \forall n \in \{1,\ldots,N\}} \sum_{n=1}^{(n)} R^{(n)}\\
    &\text{subject to\;\;} \forall n \in \{1,\ldots,N\}:\\
    &\cos^{-1}\left(\frac{\text{tr}\left(\boldsymbol{O}^{(n)} \left(\boldsymbol{O}^{n+1}\right)^{-1}\right)}{2}\right) \leq \Gamma_D \times T_s \label{eqn:opt0_cond0}\\
    &\left\| \left( x_D^{n+1} - x_D^{(n)}, y_D^{n+1} -  y_D^{(n)}, z_D^{n+1} - z_D^{(n)} \right) \right\| \leq v_D \times T_s \label{eqn:opt0_cond1}\\
    &x_{\min} \leq x^{(n)} \leq x_{\max} \label{eqn:opt0_cond2}\\
    &y_{\min} \leq y^{(n)} \leq y_{\max} \label{eqn:opt0_cond3}\\
    &z_{\min} \leq z^{(n)} \leq z_{\max} \label{eqn:opt0_cond4}
\end{align}
\end{subequations}

The goal, as defined in \eqref{eqn:optimization0}, is to optimize the cumulative throughput over the time span of $N T_s$. The constraints are:
\begin{itemize}
    \item \eqref{eqn:opt0_cond0} uses the matrix trace operator $\text{tr}(\cdot)$ to compute the rotation angle from the rotation matrices and restrict it to a maximum value, thereby adhering to the DRS's maximum rotational speed limit.
    \item \eqref{eqn:opt0_cond1} confirms the DRS's displacement per time step stays within its maximum translational speed.
    \item Constraints \eqref{eqn:opt0_cond2}, \eqref{eqn:opt0_cond3}, and \eqref{eqn:opt0_cond4} ensure the DRS's movement is confined within the designated 3D boundary.
\end{itemize}
Considering the highly non-convex nature of the objective in \eqref{eqn:optimization0}, particularly when the RIS operates in the far-field state and incorporates sinc functions as given in (\ref{eq:ris_ff_ploss}) and depicted in Fig.~\ref{fig:path_loss_surf}, and acknowledging the dynamic and intricate aspects of constraint \eqref{eqn:opt0_cond0}, we opt for a heuristic approach of setting the trajectory of the DRS, and a reinforcement learning scheme for controlling the orietnation of the DRS. This decision is driven by reinforcement learning's adaptability to complex, dynamic environments and its proficiency in handling non-linear and non-convex problems. Specifically, the trajectory of the DRS is obtained as a direct path towards the point in 3D space which maximizes the throughput of the V2X pair, and we employ a tabular Q-learning approach for rotating the RIS at each time step\footnote{In this preliminary work, we validate the effectiveness of our formulated Q-learning scheme in a tabular context, suited for smaller state and action spaces. Future expansions of this work will integrate deep learning techniques to harness greater benefits for larger state and action spaces}.

\section{Proposed Solution}
\label{sec:solution}
In describing the proposed solution, it's important understand the path-loss formula in \eqref{eq:ris_ff_ploss} which can be divided into two parts: the first part depends on the location and involves the elevation angles and distances from the RIS to each V2X pair, while the second part (represented as $\Psi$ in \eqref{eq:ris_ff_ploss}) is related to orientation and depends mainly on the azimuth angles.

The first part of the path-loss in \eqref{eq:ris_ff_ploss} changes with the DRS's position relative to the V2X pairs. It's proportional to the squared product of the distances, $d_1^2 d_2^2$, which is minimized when the DRS is positioned at the midpoint of the line segment between the V2X pair. This midpoint, where 
\begin{equation}\label{eq:equal_d}
    d_1 = d_2 = d,
\end{equation}is also where the lowest path-loss is achieved for a given height $h_D$ of the DRS, as shown in \cite{Tang2021}. Moreover, the path-loss is inversely proportional to the RIS antennas' gains, which depend on the elevation angle, $\theta$, as shown in \eqref{eq:ris_radiation}. When the DRS is at the midpoint, this angle $\theta$ for both ground nodes can be calculated as:
\begin{equation}\label{eq:equal_theta}
    \theta = \tan^{-1}\left(\frac{d}{h_D}\right).
\end{equation}

The second part of the formula in \eqref{eq:ris_ff_ploss} can be adjusted by changing the RIS's orientation in the xy plane, affecting the azimuth angles $\phi_r$ and $\phi_t$, for fixed elevation angles $\theta_r$ and $\theta_t$ that depend only on the DRS's 3D location. 

Building on the earlier analysis, we divide the optimization problem in \eqref{eqn:optimization0} into two sub-problems, which we solve using heuristic methods as outlined below.
\subsection{Setting the trajectory of the DRS $\boldsymbol{L}^{(n)}_{\text{D}}$}
While the optimal xy location for the DRS is identified as the midpoint between the V2X pairs, the optimal height at this midpoint remains to be determined. This is done by minimizing the left term of \eqref{eqn:optimization0} and therefore minimizing the following function:
\begin{equation}
    f(h_D) = \frac{d^2+H_D^2}{\cos^6\left(\tan^{-1}\left(\frac{d}{h_D}\right)\right)},
\end{equation}
where \eqref{eq:equal_d} and \eqref{eq:equal_theta} have been applied to simplify the equation to its right-hand side form. To find the optimal value of $h_D$, we can employ quasi-Newton methods like L-BFGS-B, which allows for bounded search space, ensuring compliance with condition \eqref{eqn:opt0_cond4}. In our research, we use L-BFGS-B to solve for $h_{D, \text{opt}}$.

Given the optimal location
\begin{equation}\label{eq:opt_loc}
    \boldsymbol{L}^{(n)}_{\text{opt}} = \left(\frac{x^{(n)}_1+x^{(n)}_2}{2}, \frac{y^{(n)}_1+y^{(n)}_2}{2}, h_{D,\text{opt}}\right),
\end{equation}
we define the DRS's next position at each time step, moving towards $L_{\text{opt}}$, as:
\begin{equation}\label{eq:loc_update}
    \boldsymbol{L}^{n+1} = \boldsymbol{L}^{(n)} + v_D T_s \frac{\boldsymbol{L}^{(n)}_{\text{opt}} - \boldsymbol{L}^{(n)}}{\left|\left|\boldsymbol{L}^{(n)}_{\text{opt}} - \boldsymbol{L}^{(n)}\right|\right|^2}.
\end{equation}
This approach ensures that the DRS reaches the optimal position in the shortest possible time.

\subsection{Managing the DRS's Orientation $\boldsymbol{O}^{(n)}$}

The RIS path-loss on the route taken by the DRS to arrive at the optimal midpoint is significantly influenced by the sinc functions in the right-hand side of \eqref{eq:ris_ff_ploss}, which are directly related to the azimuth angles at each time step, as clearly shown in \ref{fig:path_loss_surf}. Considering the complex and highly non-convex nature of this term, characterized by trigonometric functions in both the numerator and denominator, and the dynamic aspects of the problem, we choose to employ reinforcement learning for determining the RIS's orientation at each time step.

\subsection{$Q$-learning}
In the RL framework, at each time step $nT_s$, the agent observes the environment state $\boldsymbol s^{(n)}$ and selects an action $\boldsymbol a^{(n)}$ based on that observation. In our problem, the agent is the DRS, and its possible actions include rotating right or left with possible speeds of $\Gamma_D$, $0.75\; \Gamma_D$, $0.5\; \Gamma_D$, or $0.25\; \Gamma_D$.
At each time step $nT_s$, the environment state observed by the agent given its trajectory includes:
\begin{enumerate}
    \item Next azimuth angles $\phi^{(n+1)}_r$ and $\phi^{(n+1)}_t$,
    \item Next elevation angles $\theta^{(n+1)}_r$ and $\theta^{(n+1)}_t$,
\end{enumerate}
This can be expressed as:
\begin{equation}\label{eqn:stateDefinition}
    \boldsymbol s^{(n)} = \{\phi^{(n+1)}_r, \phi^{(n+1)}_t, \theta^{(n+1)}_r, \theta^{(n+1)}_t\}.
\end{equation}
The quantization of elevation angles is done to achieve finer granularity near 0, while azimuth angles are quantized to have greater granularity around $\pi$ and $-\pi$. The cardinalities hyperparameters resulting from this quantization for elevation and azimuth angles are denoted as $\mathcal{C_{\theta}}$ and $\mathcal{C_{\phi}}$, respectively.
After an action is chosen, the environment state changes and a reward $r^{(n)}$ is given, which guides the agent's future actions \cite{Sutton1998}. The policy $\pi(\boldsymbol s^{(n)})$ maps the observed state to the action to be taken. In our work, the reward is defined as the negative scaled path-loss from \eqref{eq:ris_ff_ploss} at each time step:
\begin{equation}\label{eqn:reward}
    r^{(n)} = -10 \times (PL^{(n)}_{f-f} - PL^{(n-1)}_{f-f}).
\end{equation}
This reward helps the agent learn and update the Q table during simulations. The agent can then be deployed in real-world scenarios, using the Sim-to-Real approach \cite{sim_to_real}.

For each state-action pair, the function $Q(\boldsymbol s^{(n)}, \boldsymbol a^{(n)})$ maps to the cumulative discounted reward, defined as $Q^\pi (\boldsymbol s^{(n)}, \boldsymbol a^{(n)}) = \mathbb{E} [R^{(n)} | \boldsymbol s^{(n)}, \boldsymbol a^{(n)}]$ where $R^{(n)}$ is the sum of future discounted rewards $R^{(n)} (\boldsymbol s^{(n)}, \boldsymbol a^{(n)}) = \sum^{\infty}_{n=1} \gamma^{(n)} r^{(n)}$, assuming future actions follow the policy $\pi(\cdot)$. The discount factor $\gamma$ in [0, 1] balances the importance of immediate versus future rewards. When an action $\boldsymbol a^{(n)}$ is taken in state $\boldsymbol s^{(n)}$, leading to reward $r^{(n)}$ and new state $s^{(n+1)}$, the Q value for the selected state-action pair is updated as:
\begin{multline}\label{eqn:qUpdate}
    Q^\text{new} (\boldsymbol s^{(n)}, \boldsymbol a^{(n)}) = Q (\boldsymbol s^{(n)}, \boldsymbol a^{(n)}) + \\ \alpha \left[r^{(n)} + \gamma \max_a Q(s^{(n+1)}, a) - Q (\boldsymbol s^{(n)}, \boldsymbol a^{(n)})\right],
\end{multline}
where $\alpha$ is the learning rate. The agent usually follows an $\varepsilon$-greedy policy where it selects the action that maximizes the Q value with probability $1-\varepsilon$ or a random action with probability $\varepsilon$. To speed up exploration of all states during exploration, in our implementation we favor selecting actions that were not selected before at each state. As for the learning rate, we adopt a decaying rate approach where 
\begin{equation}
    \alpha = \frac{1}{n^{\frac{2}{14}}}.
\end{equation}

Finally, to avoid maximization bias in Q-learning, double Q-learning is proposed, using two Q tables, $Q_1$ and $Q_2$. The action at a given state $\boldsymbol s^{(n)}$ is chosen to maximize $Q = Q_1 (\boldsymbol s^{(n)}, \boldsymbol a^{(n)}) + Q_2 (\boldsymbol s^{(n)}, \boldsymbol a^{(n)})$. For updates, one table is randomly selected at each epoch:
\begin{multline}\label{eqn:q1update}
    Q_1 (\boldsymbol s^{(n)}, \boldsymbol a^{(n)}) = Q_1 (\boldsymbol s^{(n)}, \boldsymbol a^{(n)}) +  \alpha \bigg[r^{(n)} +\\ \gamma Q_2\left(s^{(n+1)}, \argmax_a Q_1\left(\boldsymbol s^{(n+1)}, \boldsymbol a^{(n)}\right)\right) - Q_1(\boldsymbol s^{(n)}, \boldsymbol a^{(n)})\bigg]
\end{multline}
if $Q_1$ is chosen, and
\begin{multline}\label{eqn:q2update}
    Q_2 (\boldsymbol s^{(n)}, \boldsymbol a^{(n)}) = Q_2 (\boldsymbol s^{(n)}, \boldsymbol a^{(n)}) +   \alpha \bigg[r^{(n)} +\\ \gamma Q_1\left(s^{(n+1)}, \argmax_a Q_2\left(\boldsymbol s_{n+1}, \boldsymbol a^{(n)}\right)\right) - Q_2(\boldsymbol s^{(n)}, \boldsymbol a^{(n)})\bigg]
\end{multline}
if, otherwise, $Q_2$ is chosen.

At each time step, if not already serving a V2X pair, the DRS selects the active pair (i.e., with established direct link) with the closest $\boldsymbol{L}_{\text{opt}}$. That is, 
\begin{equation}\label{eq:min_dist_pair}
    k=\underset{k'}{\mathrm{\argmin}}\;\boldsymbol{L}^{(n,k')}_{\text{opt}}.
\end{equation} Then, given the distance between the pair, the optimal height of the DRS is obtained by minimizing \eqref{eq:equal_d}, and obtaining the optimal location \eqref{eq:opt_loc} at each time step given the xy locations of the V2X pairs which is used to update the DRS location according to \eqref{eq:loc_update}. While traveling towards the locations, the DRS observes the states of the environment at each time step and selects the action that maximizes the expected reward with a probability of $1-\varepsilon$.

\begin{algorithm}[H]
 \caption{DRS Trajectory and Orientation Control}
 \begin{algorithmic}[1]\label{alg}
 \renewcommand{\algorithmicrequire}{\textbf{Input:}}
 \renewcommand{\algorithmicensure}{\textbf{Output:}}
 \REQUIRE  $\gamma$, $\varepsilon$, $\boldsymbol{L}_{1}^{(n)}(t)$, $\boldsymbol{L}_{2}^{(n)}(t)$, $\boldsymbol{L}_{D}^{(n)}(t)$ $\forall n \in \{1,\ldots,N\}$
 \\ \textit{Initialisation} : initialize all values in $Q_1$ and $Q_2$ to zero
  \WHILE {An active V2X pair exists}
  \STATE $n = n + 1$
  \STATE obtain optimal 3D location from \eqref{eq:opt_loc} after solving for $h_{D,\text{opt}}$ by minimizing \eqref{eq:equal_d} for each pair \footnote{To reduce computation, only pairs with distance on the xy plane below a threshold can be considered}, and select the pair according to \eqref{eq:min_dist_pair}.
  \STATE $\boldsymbol s^{(n)} \gets$ state parameters as defined in \eqref{eqn:stateDefinition}
  \STATE With a probability of $1 - \varepsilon$ select action $a^{(n)}$ that maximizes $Q_1(\boldsymbol s^{(n)}, \boldsymbol a^{(n)}) + Q_2(\boldsymbol s^{(n)}, \boldsymbol a^{(n)})$, else
  select a random action
  \STATE Update DRS location according to \eqref{eq:loc_update}, and orientation according to selected action for $T_\text{s}$
  \STATE After $T_\text{s}$ observe reward as defined in \eqref{eqn:reward}
  \STATE With $0.5$ probability update $Q_1$ table according to \eqref{eqn:q1update}, else update $Q_2$ according to \eqref{eqn:q2update} 
  \ENDWHILE
 \end{algorithmic}
 \end{algorithm}

\section{Simulation results}
\label{sec:simres}
\subsection{Simulation setup}
As illustrated in Fig.~\ref{fig:scenario}, our scenario assumes two parallel lanes with opposite directions. The arrival times of vehicles entering the lanes are generated according to an exponential distribution with a rate parameter $\lambda_{\text{arrival}}$, and the number of V2I and V2V communication events starting at time step $n$ is Poisson distributed with densities $\lambda_{\text{v2i}}$ and $\lambda_{\text{v2v}}$. The lanes are situated parallel to the y-axis at $x=x_{\min}=0$~m and $x=x_{\max}=500$~m, and they stretch across the y axis between $y=y_{\min}=0$~m and $y=x_{\max}=5000$~m. We assume that the DRS can fly as low as $z_{\min} = 50$~m and $z_{\max} = 600$~m. Other simulation parameters are summarized in Table~\ref{tab:params}.


\begin{table}[htbp]
\begin{center}
\caption{Summary of Key Simulation Parameters}
\label{tab:params}
\begin{tabular}{||  p{5.8cm} | p{2.3cm} ||}
\hline  
\textbf{Parameter} & \textbf{Value} \\
\hline
\hline  
Antenna Gains ($G_t$, $G_r$, $G$) & 9.03 dB, 0 dB, 0 dB \\
\hline 
RIS Size ($M$, $N$), Unit Cell Size ($d_x$, $d_y$) & 100, 102, 0.01 m \\
\hline  
Reflection Coefficient ($A$), Carrier Frequency ($f_c$) & 0.9, 5 GHz \\
\hline  
Link Effectiveness ($\eta$), Effective Bandwidth ($B_{\text{eff}}$) & 0.82, 17.472 MHz \\
\hline  
V2X Power ($P_t$), Noise Power per Band ($\sigma_n^2$) & 200 mW, -131.27 dBm \\
\hline  
Time Step ($T_s$), DRS Speeds ($v_D$, $\Gamma_D$), Vehicle Speed ($v_t$) & 0.5 s, 15 m/s, 0.349 rad/s, 10 m/s \\
\hline  
Exploration Rate ($\varepsilon$), Angle Cardinalities ($\mathcal{C_{\theta}}$, $\mathcal{C_{\phi}}$) & 0.2, 100, 100 \\
\hline 
\end{tabular}
\end{center}
\end{table}


\subsection{Evaluation results and discussion}
Our simulation, spanning $6 \times 10^6$ time steps, involved the DRS serving a selected V2X pair at each step. An episode is defined from the start of serving a new V2X pair to the commencement of service to another pair. The path loss for the direct links are obtained according the models mentioned in Sec.\ref{sec:system_model}, where for V2V pairs, the presence of other vehicles between them is considered accounting fore these intermediate vehicles' height which affects the direct transmission.

\begin{figure}
      \centering
\includegraphics[width=0.45\textwidth]{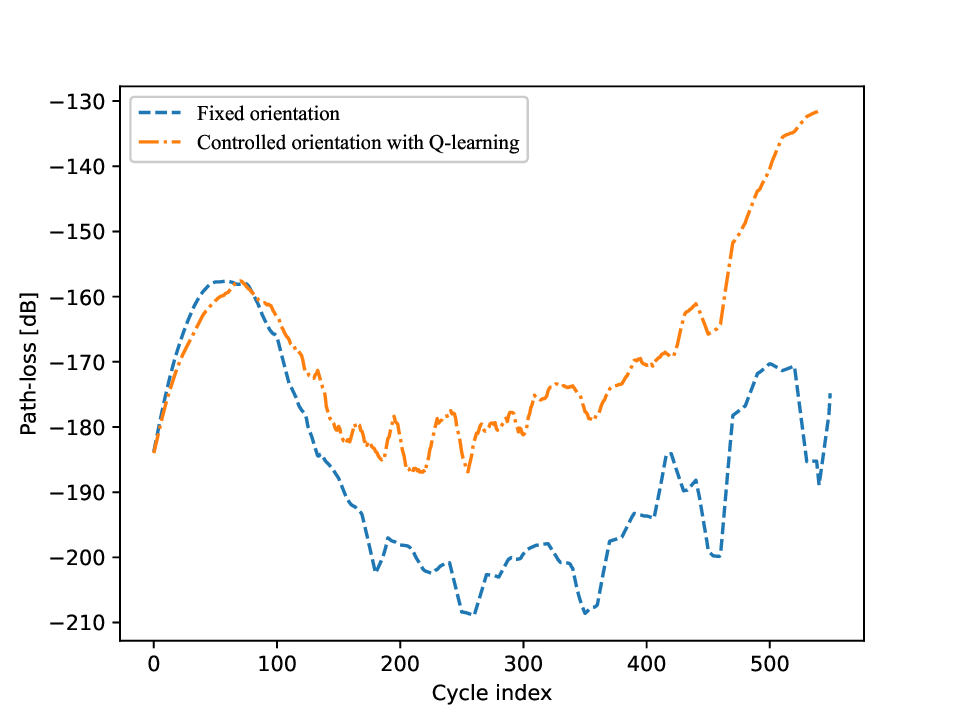}
  \caption{
Average path-loss per cycle during the connectivity duration of V2X pairs, ending as the DRS reaches its optimal location. The use of the Q-learning method clearly improves path loss reduction.}
  \label{fig:reward}
\end{figure}

Figure~\ref{fig:reward} displays the average rewards per cycle for varying episode lengths, which depend on the V2X pairs' relative positions along the path and their duration within the simulation's boundaries. The length of an episode can be short if the DRS selects a V2X pair near the simulation border, nearing the end of its presence in the simulation. Such episodes affect the average reward shown in the figure for the initial cycles. This scenario leads to higher rewards when the DRS is already close to that border (which is probable since the DRS tracks pairs until the end of their simulation to either border), which entails that the DRS quickly reaches pairs already close to its position. However, the results show marked improvement in longer episodes, highlighting the effectiveness of our Q-learning strategy, which is based on tabular learning with a limited state space. The optimization of the RIS's orientation is particularly crucial when the DRS's speed is insufficient to reach the optimal location swiftly, especially when dealing with high-speed vehicles. As the DRS nears to the optimal location, the impact of the RIS's orientation lessens, reducing the significance of azimuth and elevation angles in \eqref{eq:ris_ff_ploss}.

\begin{figure}
      \centering
\includegraphics[width=0.4\textwidth]{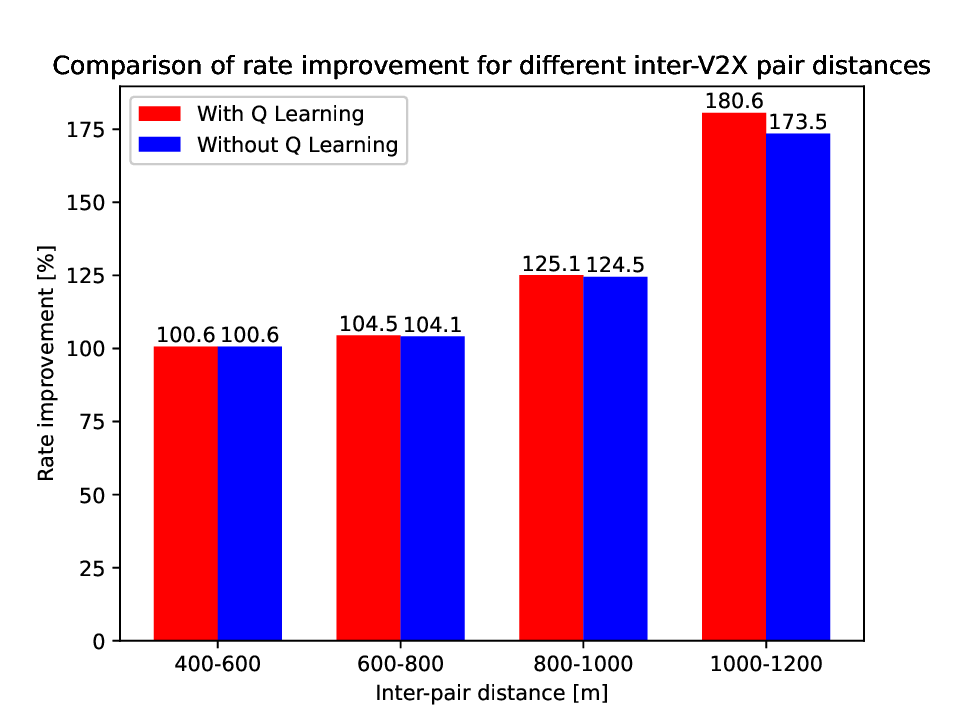}
  \caption{
Average improvement in rate for various allowed inter-v2x pairs distances during testing (i.e., $\varepsilon=0$). As the distance increases the direct link quality reduces and the RIS LOS link is more important.}
  \label{fig:reward_improvement}
\end{figure}

The second figure, Fig. \ref{fig:reward_improvement}, illustrates the performance of our proposed solution across various inter-V2X pair distance ranges during testing in terms of average path-loss. As the distance between pairs increases, the benefit of employing the DRS becomes more pronounced, providing a virtual line-of-sight path that circumvents potential blockages like cars or obstacles. This enhancement not only extends the viable distance range for V2X pair communication but also contributes to the overall efficiency of the V2X communication system. It facilitates faster payload transmission for certain pairs, potentially freeing up resources for better scheduling among other pairs. Additionally, the figure highlights the marked improvement achieved by optimizing the DRS's orientation throughout its trajectory.
\section{Conclusions}
\label{sec:concl}
In this paper we investigated the application of DRS equipped with RIS to support V2X communications. Due to the complexity and non-convexity of the investigated problem, we proposed a heuristic reinforcement learning-based approach to optimize the path and orientation of DRS. The optimal drone location is found analytically using geometrical relations, while the orientation is learned via the Q-learning strategy. 
The proposed approach's effectiveness, confirmed by simulations, improved communication rates for the selected V2X pairs. As only a simple Q-learning approach was applied, we have limited the optimization to a single selected pair of vehicles to avoid potential conflicts and ensure algorithm convergence. Future efforts with deep learning will address interference issues and enable simultaneous support for multiple V2X pairs with one DRS, along with optimizing the selection of V2X pairs in dynamic environments.

\bibliography{IEEEabrv,percom_biblio}
\bibliographystyle{IEEEtran}

\end{document}